\documentstyle[12pt]{article}
\catcode`\@=11

\def\section{\@startsection{section}{1}{\z@}
  {3.5ex plus 1.0ex minus 0.2ex}{2.3ex plus .2ex}{\normalsize}}
\def\subsection{\@startsection{subsection}{2}{\z@}
  {3.25ex plus 1.0ex minus 0.2ex}{1.5ex plus 0.2ex}{\normalsize\bf}}
\def\subsubsection{\@startsection{subsubsection}{3}{\z@}
  {3.25ex plus 1.0ex minus 0.2ex}{1.5ex plus 0.2ex}{\normalsize\bf}}

\@addtoreset{equation}{section}

\catcode`\@=12

\newcommand{\skipline}{\vspace{\baselineskip}}

\def\fun#1#2{\lower3.6pt\vbox{\baselineskip0pt\lineskip.9pt
  \ialign{$\mathsurround=0pt#1\hfil##\hfil$\crcr#2\crcr\sim\crcr}}}

\begin{document}
\begin{titlepage}
\begin{center}
Infinite Volume, Continuum Limit of Valence Approximation Hadron Masses
\end{center}
\skipline
\skipline
\begin{center}
F. Butler, H. Chen, A. Vaccarino, J. Sexton \footnote{permanent address
Dept. of Mathematics, Trinity College, Dublin 2, Ireland} and D.
Weingarten\footnote{talk presented by D. Weingarten at Lattice 92,
Amsterdam, The Netherlands, Sept. 15-19, 1992}\\
IBM Research \\
P.O. Box 218, Yorktown Heights, NY 10598
\end{center}
\skipline
\skipline
\begin{center}
ABSTRACT
\end{center}
\skipline
\begin{quotation}
We obtain estimates of several hadron mass ratios, for Wilson quarks in
the valence (quenched) approximation, extrapolated to physical quark
mass, infinite volume and zero lattice spacing.
\end{quotation}
\skipline
\skipline
\end{titlepage}

In the present article, we evaluate the pion, rho, proton and delta
baryon masses using Wilson quarks in the valence (quenched)
approximation for a range of different choices of quark mass, lattice
volume, and lattice spacing. From these results we obtain estimates of
the infinite volume continuum limits of eight different hadron mass
ratios, evaluated at physical quark masses. Seven of these ratios differ
from experiment by less than 6\% which, in each case, is less than a
factor of 1.5 multiplied by the corresponding statistical uncertainty.
{}From the lattice gauge coupling constant, $g_{lat}$, we determine the
continuum coupling constant, $g^{(0)}_{\overline{ms}}$
\cite{Fermilab}.  Values of the rho mass in lattice units depend on
$g^{(0)}_{\overline{ms}}$ as predicted by asymptotic scaling, for a
$\Lambda^{(0)}_{\overline{ms}}$ within one standard deviation of the
continuum limit value found in Ref.~\cite{Fermilab}.

The motivation for this work is both to compare the valence
approximation with experiment and to develop technology which will be
useful in extrapolating results of the full theory to physical quark
mass, infinite volume, and zero lattice spacing.

The calculations reported here were done on the GF11 parallel computer
at IBM Research \cite{Weingar90} and took approximately one year to
complete.  GF11 was used in configurations ranging from 384 to 480
processors, with sustained speeds ranging from 5 Gflops to 7 Gflops.
With the present set of improved algorithms and 480 processors, these
calculations could be repeated in less than 4 months.

Hadron propagators were evaluated at $\beta = 5.7$ and $k = 0.1400 -
0.1675$ using 2349 configurations on a lattice $8^3 \times 32$, 219
configurations on a lattice $16^3 \times 32$ and 92 configurations on a
lattice $24^3 \times 32$. At $\beta = 5.93$ and $k = 0.1543 - 0.1581$,
217 configurations were used on a lattice $24^3 \times 36$. At $\beta =
6.17$ and $k = 0.1500 - 0.1532$, 219 configurations were used on a
lattice $30 \times 32^2 \times 40$.  Gauge configurations were generated
by a version of the Cabbibo-Marinari-Okawa algorithm adapted for
parallel computers \cite{CMO}.  The number of updating sweeps between
successive configurations ranged from 1000 for the $8^3 \times 32$
lattice at $\beta$ of 5.7 to 6000 on the $30 \times 32^2 \times 40$
lattice at $\beta$ of 6.17.  A variety of correlation tests showed all
of the configurations on which propagators were evaluated were
statistically independent.

For the $8^3 \times 32$ lattice at $\beta$ of 5.7 we used point sources
and sinks in the hadron propagators. For all other lattices and $\beta$,
gaussian extended sources and both point sinks and gaussian
sinks\cite{Smear} were chosen. The mean squared radius of the gaussian
source in all cases was 6 in lattice units.  The mean squared radii of
the gaussian sinks in lattice units were $3d^2/2$, for $d$ of 1 to 4.
On the lattice $24^3 \times 32$ at $\beta$ of 5.7, 8 independent
gaussian sources were placed on the source hyperplane\cite{Butler}, each
multiplied by a random cube root of 1 to cancel cross terms between the
propagation of different sources. Random cube roots of 1 cause cross
term cancellation both for meson and baryon propagators.

Each gauge configuration was first transformed to a completely fixed
axial gauge and then to lattice Coulomb gauge. Since the axial gauge is
uniquely defined and has no Gribov copies, the final Coulomb gauge
transformation selects a uniquely specified Gribov copy which would not
be changed by any gauge transformation of the original configuration.

Quark propagators were constructed using the conjugate gradient
algorithm for the $8^3 \times 32$ lattice at $\beta$ of 5.7, using
red-black preconditioned conjugate gradient for the other lattices at
$\beta$ of 5.7 and 5.93, and using a red-black preconditioned minimum
residual algorithm at $\beta$ of 6.17. For conjugate gradient, at the
largest values of the hopping constant k, we found that red-black
preconditioning decreased the computer time required for each inversion
by very close to a factor of 3. At the largest $\beta$ the minimum
residual algorithm gave an additional improvement by a factor of 2 over
conjugate gradient.  For each lattice and $\beta$, we found no
configurations on which the inversion algorithm chosen failed to
converge.  The convergence criterion used in all cases was tuned to be
equivalent to the requirement that effective pion, rho, nucleon and
delta masses evaluated between successive pairs of time slices must be
within 0.2\% of their values obtained on propatagors run to machine
precision.

Hadron masses were then determined by fits to hadron propagators
constructed from the quark propagators.  In all cases the pion mass was
extracted by fitting propagator data for a point sink, on an interval of
hyperplanes sufficiently far from the propagator's source, to
$Z\{exp(-mt)+exp[-m(N-t)]\}$ for hyperplane $t$ on a lattice with
time-direction periodicity $N$. The best $Z$ and $m$ were found by
minimizing the fit's $\chi^2$ determined from the full statistical
correlation matrix among the set of fitted hyperplanes. The range to be
fit was first narrowed down by looking for an approximate plateau in the
large $t$ tail of graphs of the naive pion mass found by fitting only
succesive pairs of hyperplanes to $Z \{ exp( -m t) + exp[ -m (N-t)]\}$.
The fitting program then chose an optimal fitting range within this
region by trying all ranges including at least three hyperplanes and
selecting the range yielding the smallest value of $\chi^2$ per degree
of freedom. A corresponding procedure was used to determine the rho,
nucleon and delta baryon masses for all values of $k$ on the lattice
$8^3 \times 32$ and for all but the three largest values of $k$ on the
other four lattices.  For the nucleon and delta baryon the fitting
function used was $Z exp(-mt)$.

At the largest three $k$ values, the rho mass was found by
simultaneously fitting the propagators for a point sink and for gaussian
sinks with mean squared radius of $3d^2/2$, for $d$ of 1 and 2.  The
fitting function was taken to be $Z_d\{exp(-mt)+exp[-m(N-t)]\}$ with
field strength renormalization constants $Z_0, Z_1, Z_2,$ depending on
the sink but mass $m$ independent of the sink. The optimal fitting
parameters and fitting range were found by minimizing $\chi^2$
determined by the full correlation matrix among the complete set of
fitted hyperplanes and sink sizes. Nucleon and delta baryon masses were
determined similarly using the fitting function $Z_d exp(-mt)$.
Simultaneous fits to several sinks at once provides an unbiased
resolution of the small differences between mass results obtained from
fits to different individual sinks and give statistical uncertainties
which are frequently smaller by a factor ranging from 0.85 to 0.9 than
the smallest statistical uncertainty obtained by fitting a propagator
for a single sink size.

The statistical errors for all fits were determined by the bootstrap
method.  Generating 100 bootstrap ensembles in all cases appeared to be
sufficient to give stable values for statistical errors.

Comparing hadron masses in lattice units between the $8^3 \times 32$ and
$16^3 \times 32$ lattices at $\beta$ of 5.7 for $k$ up to 0.1650 showed
no statistically significant differences. Comparing $16^3 \times 32$ and
$24^3 \times 32$ for $k$ up to 0.1675 showed no statistically
significant differences in the pion mass for fixed $k$. Overall, the
pion mass at fixed $k$ varied by less than 1.2 $\pm$ 1.6 \% as the
lattice size was changed. For the rho, nucleon and delta baryon, some
differences were found comparing the $16^3 \times 32$ and $24^3 \times
32$ lattices. At $k$ of 0.16625, the nucleon mass fell by 4.4 $\pm$ 1.7
\% from $16^3 \times 32$ to $24^3 \times 32$. At $k$ of 0.1675, the rho
fell by 3.4 $\pm$ 1.4 \%, the nucleon fell by 4.6 $\pm$ 2.2 \% and the
delta baryon fell by 4.7 $\pm$ 2.6 \% from $16^3 \times 32$ to $24^3
\times 32$.  It appears quite likely that for the range of $k$, $\beta$
and lattice volume we have examined, the errors in valence approximation
hadron masses due to calculation in a finite volume $L^3$ are bounded by
an expression of the form $C exp( - L/R)$, with a coefficient $R$ of the
order of the radius of a the hadron's wave function. At $\beta$ of 5.7
for the $k$ we considered, $R$ is thus typically 3 lattice units. Thus
we expect that the differences we have found between masses on a $16^3$
volume and those on a $24^3$ volume are nearly equal to the differences
between $16^3$ and true infinite volume limiting values.

At the largest $k$ on each lattice, except $8^3 \times 32$, the ratio
$m_{\pi} / m_{\rho}$ was close to 0.5. Thus to produce mass predictions
for hadrons containing only ordinary quarks our data had to be
extrapolated to still larger $k$. To do this we first determined the
$k_{crit}$ at which $m_{\pi}$ becomes 0.  As expected from a naive
application of PCAC to Wilson fermions, we found $(m_{\pi} a)^2$ to be
close to a linear function of $1/k$ over the entire range of $k$
considered on each lattice.  Fits of $(m_{\pi} a)^2$ to linear functions
of $1/k$ at the three highest $k$ on the lattices $16^3 \times 32$,
$24^3 \times 32$, $24^3 \times 36$ and $30 \times 32^2 \times 40$ gave
an averaged $\chi^2$ per degree of freedom of 1.20.  These fits were
then used to determine $k_{crit}$ for each lattice and $\beta$.
Defining the quark mass in lattice units $m_q a$ to be $1/(2 k) - 1/(2
k_{crit})$, we found $m_{\rho} a$, $m_N a$ and $m_{\Delta} a$ to be
nearly linear functions of $m_q a$ over the entire range of $k$
considered on each lattice. Figure~\ref{fig:nuc_rho} shows the nucleon
mass $m_N$ and rho mass $m_{\rho}$ as functions of the quark mass $m_q$
in comparision to linear fits to the 3 smallest values of $m_q$. The
hadron masses are shown in units of the physical value of $m_{\rho}$,
given by $m_{\rho}$ evaluated at the ``normal'' quark mass $m_n$ which
produces the physical value of $m_{\pi}/m_{\rho}$. The quark mass $m_q$
in Figure~\ref{fig:nuc_rho} is measured in units of the strange quark
mass $m_s$, the determination of which will be discussed below. The fits
shown in Figure~\ref{fig:nuc_rho} appear to be reasonably good and
should form a fairly reliable method for extrapolations down to light
quark masses.  Fits comparable to those shown were obtained for the
nucleon, rho and delta baryon on all the lattices we considered except
$8^3 \times 32$.

\begin{figure}
\vspace{5 in}
\caption{For $16^3 \times 32$ at $\beta$ of 5.7, $m_N$ and $m_{\rho}$ in
units of the physical $m_{\rho}$ as a function of the quark mass $m_q$
in units of $m_s$}
\label{fig:nuc_rho}
\end{figure}

A further test of our extrapolation method is given by a version of the
Gell Mann-Okubo mass formula. For a rho composed of quarks with masses
$m_1$ and $m_2$, assume, as suggested by Figure~\ref{fig:nuc_rho}, that
$m_{\rho} = \alpha_1 m_1 + \alpha_2 m_2 + \beta$.  It is then easy to
show we must have $\alpha_1 = \alpha_2$.  Then the k-star, equivalent to
a rho with $m_1 = m_s$ and $m_2 = m_n$, will have the
same mass as a rho composed of a single type of quark with $m_1 = m_2 =
(m_s + m_n)/2$.  In the valence approximation the phi is equivalent to a
rho with $m_1 = m_2 = m_s$. By a further application of the linearity
relation of Figure~\ref{fig:nuc_rho}, we can then extrapolate the rho
mass from the masses of k-star and phi. The extrapolated rho mass
obtained in this way from observed k-star and phi masses lies below the
observed rho mass by 0.53 \%.  A similar extrapolation can be made
to determine the nucleon mass from the observed masses of its strange
partners, and to determine the delta baryon mass from its strange
partners.  The nucleon mass found this way is 1.38 \% too large, and the
delta baryon mass is 0.81 \% too large.

The relations discussed in the preceding paragraph, conversely, permit
strange hadron masses to be constructed from the masses we have
calculated for hadrons composed of a single species of heavy quark.
Fitting the pseudoscalar kaon to the pion mass at a quark mass of $(m_s
+ m_n)/2$ gives the value for $m_s$ mentioned earlier. With $m_s$ and
$m_n$ thus completely fixed, predictions for eight different hadron mass
combinations, measured in units of the physical rho mass, follow from
our data with no additional free parameters.

Finally, to obtain predictions for comparision with experiment we have
extrapolated our results to 0 lattice spacing. For Wilson fermions
we expect the leading lattice spacing
dependence in mass ratios to be linear in a. A linear fit of our
data for $m_N / m_{\rho}$ to the lattice spacing, measured in units of
the physical rho mass, is shown in Figure~\ref{fig:nuc_extrap}. The
vertical bar at 0 lattice spacing is the extrapolated prediction's
uncertainty, determined by the bootstrap method. The dot at 0 lattice
spacing shows the observed physical value of $m_N / m_{\rho}$.
The three data points in Figure~\ref{fig:nuc_extrap} are for the
lattices $16^3 \times 32$, $24^3 \times 36$ and $30 \times 32^2 \times
40$.  The values of $\beta$ for these lattices were chosen so that the
physical volume in each case is nearly the same. The lattice period L,
measured in units of the physical rho mass, $m_{\rho} L$ is,
respectively, 9.00 $\pm$ 0.13, 9.15 $\pm$ 0.19 and 8.76 $\pm$ 0.13.

\begin{figure}
\vspace{5 in}
\caption{$m_N/m_{\rho}$ as a function of the lattice spacing, $m_{\rho} a$,
extrapolated to 0 lattice spacing.}
\label{fig:nuc_extrap}
\end{figure}

Linear extrapolations to 0 lattice spacing for eight different hadron
mass ratios, and the corresponding observed values, are shown in
Table~\ref{tab:results}. Values of $\Lambda^{(0)}_{\overline{ms}}$,
found following Ref.~\cite{Fermilab}, give
$\Lambda^{(0)}_{\overline{ms}} / m_{\rho}$ which vary by only one
standard deviation over the three lattices used for extrapolation. Thus
the rho mass follows asymptotic scaling in $g^{(0)}_{\overline{ms}}$.
For the observed value of $\Lambda^{(0)}_{\overline{ms}}$ we have
inserted the calculated result of Ref.~\cite{Fermilab}.  Seven of the
eight predictions differ from experiment by less than 6\% and less than
1.5 multiplied by the average of the upper and lower error bars.

\begin{table}
\begin{center}
\begin{tabular}{|l|r|r|}     \hline
 ratio & calculated & observed \\ \hline

 $m_{K^*} / m_{\rho}$ & 1.150 + 0.008 & 1.161 \\
                            &       - 0.012 &       \\ \hline
 $m_{\Phi} / m_{\rho}$ & 1.300 + 0.015 &  1.327 \\
                             &       - 0.024 &        \\ \hline
 $m_N / m_{\rho}$ & 1.284 + 0.071 & 1.222 \\
                          &       - 0.065 &       \\ \hline
 $(m_{\Xi} + m_{\Sigma}- m_N) / m_{\rho}$ & 1.869 + 0.030 & 2.046 \\
        &       - 0.063 &       \\ \hline
 $m_{\Delta} / m_{\rho}$ & 1.627 + 0.051  &  1.604 \\
                               &       - 0.092  &        \\ \hline
 $m_{\Sigma^*} / m_{\rho}$ & 1.816 + 0.029 & 1.803 \\
                                 &       - 0.074 &       \\ \hline
 $m_{\Xi^*} / m_{\rho}$ & 2.017 + 0.024  &  1.994 \\
                              &       - 0.078  &        \\ \hline
 $m_{\Omega} / m_{\rho}$ & 2.210 + 0.047 & 2.176 \\
                               &       - 0.068 &       \\ \hline
 $\Lambda^{(0)}_{\overline{ms}} / m_{\rho}$ & 0.3047 + 0.0050 & 0.305 + 0.018
\\
                                 &        - 0.0109 &        - 0.018 \\ \hline
\end{tabular}
\caption{The continuum limits of hadron mass ratios in comparison
to their observed values.}
\label{tab:results}
\end{center}
\end{table}

The authors are grateful to Chi-Chai Huang, of Compunetix Inc., for his
contributions to bringing GF11 up to full power and keeping it in
operation during the course of this work.

\end{document}